# Assessing Excel VBA Suitability for Monte Carlo Simulation

**Alexei Botchkarev**
GS Research & Consulting
Adjunct Professor, Ryerson University
Toronto, Ontario, Canada
alex.bot@gsrc.ca    www.gsrc.ca

**Abstract**

Monte Carlo (MC) simulation includes a wide range of stochastic techniques used to quantitatively evaluate the behavior of complex systems or processes. Microsoft Excel spreadsheets with Visual Basic for Applications (VBA) software is, arguably, the most commonly employed general purpose tool for MC simulation. Despite the popularity of the Excel in many industries and educational institutions, it has been repeatedly criticized for its flaws and often described as questionable, if not completely unsuitable, for statistical problems. The purpose of this study is to assess suitability of the Excel (specifically its 2010 and 2013 versions) with VBA programming as a tool for MC simulation. The results of the study indicate that Microsoft Excel (versions 2010 and 2013) is a strong Monte Carlo simulation application offering a solid framework of core simulation components including spreadsheets for data input and output, VBA development environment and summary statistics functions. This framework should be complemented with an external high-quality pseudo-random number generator added as a VBA module. A large and diverse category of Excel's incidental simulation components that includes statistical distributions, linear and non-linear regression and other statistical, engineering and business functions require execution of due diligence to determine their suitability for a specific MC project.

## 1. Introduction

Monte Carlo (MC) methods [1, 2, 3] denote a wide range of stochastic techniques based on generating probability distributions as inputs to model uncertainty and randomly sampling through multiple repeated runs (simulations) to quantitatively evaluate the characteristics and behavior of complex systems or processes. Examples of applications include sensitivity analysis, error propagation, numerical integration, etc. MC methods are widely used by scientists, engineers, mathematicians, statisticians to solve problems in engineering [4], physics [5], applied statistics [6], medicine [7], nanobiotechnology [8], economics [9, 10], finance [11], manufacturing and business [12] and many other fields.

Computer realization of the Monte Carlo method can be achieved through several approaches [10, 13].

---

First, MC models can be coded from scratch with a high-level programming language. Most commonly used languages are Fortran, C, C++, JAVA, etc. These languages have libraries of frequently used statistical functions to facilitate program development. Usually, this approach is used to develop tailor-made programs to address specific situations.

Second, commercial software packages exist that provide MC simulation environment and components to facilitate modelling and simulation, e.g. ExtendSim [14], Stata [15], Arena Simulation Software [16], gretl [17], Simulink, and general purposes statistical packages SAS, MATLAB, R, Stata, SPSS. Abu-Taieh et al [18] and Pezotta et al [19] performed comparative studies of the commercial simulation packages.

Third, spreadsheet software packages are, arguably, the most commonly used general purpose tools for MC simulation [20, 21]. Thomas Schriber rightfully dubbed spreadsheet-based MC simulation as simulation for the masses [22]. Spreadsheet simulation gained its popularity due to [23]:

- Availability. Spreadsheet software (most often Microsoft Excel) is usually available to all employees in corporations, governments and to students in educational institutions as a standard tool used for many purposes.
- User developed knowledge and skills. Most employees and students have knowledge and skills to use Excel which they developed earlier, and are current with these skills. There's no need to start learning any new specialized software packages.
- Simplicity. Simulation-related skills are easy to teach and develop.
- Intuitive visualization. Spreadsheet presentation construct allows for clear visualization of the simulation phenomena.
- Broad range of applications. As a general purpose tool, Excel can be applied to solve a wide range of problems.

Multiple papers study the use of the spreadsheets for MC simulation and apply spreadsheets to solve practical problems. Barreto and Howland studied application of MC simulation with Excel to econometrics problems [37]. Menn and Holle used Excel with VBA for health economic evaluations [38]. Kying explored Excel multivariate MC simulation as it applies to economic valuation of complex financial contracts [39]. Wang et al presented a practical approach to slope stability reliability analysis using spreadsheet MC simulations [36]. Gedam and Beaudet used spreadsheet MC simulation for predicting reliability of complex systems [45]. Dobrican proposed using Excel MC simulation to forecast demand for automotive aftermarket inventories [46]. Rozycki used Excel-based MC simulation as a capital budgeting risk management tool [35]. Wang and Cao applied spreadsheet MC simulation to geotechnical analysis [40]. Au and Wang used Excel MC simulation for engineering risk assessment [41].

Also, there is a subcluster of papers on using Excel spreadsheets as a tool to teach MC simulation concepts. Mielczrek and Zabava discussed spreadsheet MC simulation in teaching management science [25, 27]. Lee demonstrated effective use of spreadsheet simulation to teach project management [28]. Briand and Hill used Excel to teach Monte-Carlo experiments to undergraduates in an econometrics course [29]. Yin and Leon taught data resampling to students from business, accounting and economics using Excel for MC simulation [31]. Pecherska and Merkuryev used spreadsheets in teaching simulation concepts [42].

It should be noted that the studies on spreadsheet MC simulation mentioned above are focused on application of Excel to solve practical problems. The researchers accept Excel as a powerful and tested tool, and skip a step of doing due diligence in verifying suitability of applying Excel to their specific problems.

Despite the popularity of the Excel in many industries and educational institutions, it has been repeatedly criticized for its flaws: low quality random numbers generator (RNG), inaccuracies of statistical functions [43, 44]. Unfortunately, Microsoft's corrective efforts to address concerns of the statistical community were too slow, too late: some identified and well-documented Excel deficiencies were moving from one version to the next not attended. These delays aggravated criticism:

> "*Excel has been found inadequate in all three areas* [estimation, random number generation statistical distributions]… *We advise that Excel not be used for statistical calculations.*" [72]
>
> "*Persons desiring to conduct statistical analyses of data are advised not to use Excel 2003.*" [73]
>
> "*… Microsoft had taken Excel through five major revisions (Excel 5.0, 95, 97, 2000, and XP) without correcting easily remedied calculation difficulties.*" [75]
>
> "*Microsoft Excel spreadsheets have become somewhat of a standard for data storage, at least for smaller data sets. This, along with the program often being packaged with new computers, naturally encourages its use for statistical analyses. This is unfortunate, since Excel is most decidedly **not** a statistical package.*" [76]
>
> "*No statistical procedure in Excel should be used until Microsoft documents that the procedure is correct; it is not safe to assume that Microsoft Excel's statistical procedures give the correct answer. Persons who wish to conduct statistical analyses should use some other package.*" [79]
>
> "*… introductory texts on statistics warn students not to use Excel when the results matter*" [79]
>
> "*…Excel provides a number of standard statistical and graphing procedures. However, these should be approached with caution, as statisticians have found numerous errors in Excel's statistical routines and distributions. … For these reasons, we do not recommend using Excel for statistical analysis, beyond very basic descriptive statistics and getting a feel for your data.*" [81]
>
> "*Due to substantial deficiencies, Excel should not be used for statistical analysis. We should discourage students and practitioners from such use. … Friends Don't Let Friends Use Excel for Statistics!*" [82]

Critics may leave an impression that Excel MC simulation is completely inaccurate and is comparable to the Buffon's needle experiment to calculate π with 300 tosses, and could be used only when numerical results don't matter.

This paper is focused on Excel. Negative remarks presented above shouldn't be understood that other statistical tools are perfect. Excel is not the only statistical tool which was subjected to criticism. McCullough assessed three very popular statistical packages (SAS, SPSS and S-Plus) and found flaws in all three areas of evaluation: estimation, random number generation, and statistical distributions [69]. The same refers to the results of assessments of several types of econometric software packages (EViews, LIMDEP, SHAZAM and TSP) [70, 71].

Motivation for this study was to tackle an existing knowledge gap represented by two disconnected trends in the academic literature on Monte Carlo spreadsheet simulation:

- The first is devoted to the use of the Excel for MC simulation to solve practical industry-specific problems without considering the risks and limitations related to the use of the tool.
- The second (mostly expressed by the statistical community) is focused on the generic limitations of the Excel as a statistical tool and describes Excel as a questionable, if not completely unsuitable, tool for MC simulations without considering specificity of the practical problems the tool is used to solve.

We believe that both trends lack systems vision. Selection of the software tool for MC simulation shouldn't be based exclusively on the availability of the tool ("Excel is the only one we have"), neither on the assumption that the tool is appropriate because "everybody else is using it". At the same time, the tool shouldn't be discarded from a list of potential options because of generic limitations which might be irrelevant for a particular problem to be solved. Selection of the tool (and that applies to selection of any software tool for any purpose) should be based on matching the requirements dictated by the problem at hand and known software limitations.

**The purpose of this study** is to assess suitability of the Excel (specifically its 2010 and 2013 versions) with VBA programming as a tool for MC simulation.

**Research Questions:**

What is the status of the statistical capabilities of the most recent Excel versions (in view of the prior known problems)?

What errors and uncertainties are typical for the studies using MC simulation and how to deal with them?

Is it appropriate to use Excel 2010 and Excel 2013 with VBA programming for MC simulation?

What are the limitations (if any) of using Excel 2010 and Excel 2013 with VBA programming for MC simulation?

Several methodologies were used to achieve the research objectives: critical literature review, spreadsheet numerical experiment, critical thinking and inductive reasoning.

The following is out of scope of this paper:

> Excel has been around for a couple of decades. During this time, many successive versions of Excel have been rolled out (and are still being used). To avoid confusions and lengthy comparisons, we focus on the most recent versions – 2010 and later, and set earlier Excel versions out of scope.

> Excel spreadsheets can be used with add-ins – supplementive software designed to enhance certain capabilities of Excel to perform MC. Most commonly used add-ins are Crystal Ball from Oracle, @RISK from Palisade, Risk Solver add-in from Frontline Systems, etc. [51, 54]. To make analysis more straightforward, we focus on the "native" Excel and set add-ins out of scope.

Our focus is on the use of Excel as a platform for MC simulations, and Excel features that affect such application. Due to that Excel deficiencies and strengths that do not directly limit MC simulations have been left out of scope.

Excel provides a large set of functions, procedures, etc. Sometimes, Excel is criticized for not having certain capabilities. For example, Data Analysis ToolPak doesn't contain tools for calculating nonparametric test statistics [47]. Another example is missing function for the right tail probabilities for the Poisson distribution [86]. We consider these issues out of scope.

The paper is intended for academics and researchers interested in MC simulation. It can also be used by practitioners in a broad subject area for evaluating potential tools for MC simulation.

The rest of the paper is organized as follows. In Section 2, we present a literature review of the main Excel statistical components that are commonly used in the MC studies: random number generators, statistical distributions, summary statistics, linear and non-linear regression. In Section 3, we analyse typical errors encountered in MC simulation and emphasize that the errors induced by the statistical tool is only one source of a variety of errors inherent to MC modelling and simulation. The section also presents the best practices of dealing with errors and uncertainties in the MC studies through the validation and verification techniques. In Section 4, we present results of the numerical experiment. Section 5 is devoted to the discussion of the results and Section 6 offers concluding remarks.

## 2. Literature Review

Literature review has been conducted in several subject areas that present interest to most types of MC simulation: random numbers generators, univariate summary statistics, statistical distributions, linear and non-linear regression.

### 2.1. Random Number Generators

**Overview of Random Number Generators.** Random number generators or, to be precise, pseudo-random generators (PRNGs) are the key components of the MC computer simulators.

> *"An ideal random number generator would provide numbers that are uniformly distributed, uncorrelated, satisfy any statistical test of randomness, have a large period of repetition, can be changed by adjusting an initial "seed" value, are repeatable, portable, and can be generated rapidly using minimal computer memory."*[24]

PRNG should meet certain desirable quality criteria [5, 34, 48, 89, 94]:

- Good statistical properties of the output numbers.
- Large period. By definition, algorithmic PRNG output is periodic. The expectation is that the length of the period should be significant. Expectations regarding the minimum required period vary from $2^{128}$ [30], to $10^{50}$ (~$2^{166}$) [34], to $2^{200}$. Very long periods were cited (WELL1024a with period $2^{1024} − 1$, WELLRNG44497 with period $2^{44497}−1$ [13]). Sufficiency of the period is determined by the needs of application.
- Theoretical foundation. A PRNG should be based on solid mathematical foundation allowing for analytical solutions for the PRNG properties including period length.
- Repeatability. A PRNG should be able to reproduce exactly the same sequence of numbers.

- Computational efficiency. A PRNG should display high speed of generating numbers and use minimum amount of memory.
- Portability. A PRNG should be able to perform the same way on different software/hardware platforms.

There's no mandatory set of requirements that any "usable" PRNG must meet. Only in relation to a specific application certain quality criteria may be given higher or lower priority or discarded altogether. Many PRNGs have been developed using a variety of underlying mathematical algorithms and output qualities [e.g. 26, 34, 89, 94]. In case of insufficient period, two or more PRNGs can be combined to avoid regular patterns and increase period [32, 33]. Mersenne Twister algorithm has been recognized as a valid choice for MC simulation [96]. PRNGs based on this algorithm passed statistical tests and have large periods: e.g. MT11213 a period of $2^{11213}-1$ and MT19937 has a period of $2^{19937}-1$ (approximately $10^{6001}$) [34, 94]. Recently, another group of random algorithms was proposed by Panneton, L'Ecuyer and Matsumoto - Well Equidistributed Long-period Linear (WELL) [97]. It has implementations with a period of up to 44497 and alleviates some imperfections of the Mersenne Twister PRNGs.

A variety of statistical tests have been designed to evaluate statistical qualities of the PRNGs. Primarily their purpose is to detect correlation and deviations from uniformity (equidistribucity). The practical importance of statistical testing of PRNGs cannot overestimated because certain types of correlations and other stochastic imperfections can lead to large systematic errors in MC simulation [25, 92]. Most commonly accepted tests are the TestU01 [52], Diehard tests [53], or its expanded open source version Dieharder [95]. They involve large data sets and multiple algorithms. New and more comprehensive tests regularly being suggested. To a certain extent, this reflects the growing and changing understanding of the phenomenon of *randomness* by scientists. Newly developed tests may even "take out of business" PRNGs that have been used for years and were trusted. For example, that happened to the RANDU - a standard PRNG for IBM 360 and 370 series [2, 94]: unacceptable correlation was found in the output numbers. Although, arguably, almost everything we used to fly, launch and fire in the second part of the 20[th] century was based on MC simulation using RANDU. L'Ecuyer was right stating that:

> *"Of course, the quality of a generator can never be proven by any statistical test."*[32]

Another practical approach to test the quality of a generator is to use it for MC simulation for a problem that has a proven numerical theoretical solution. The result of the simulation can be compared to the known value. Obvious condition for using this approach is that the exact solution should be known and MC algorithm should be validated. With this approach, any simulation problem can become a statistical test. Coddington has shown that many widely used generators that passed certain randomness tests and were recommended in many text books and included in some commercial software products failed this "reality check" [24, 48]. It should be noted that passing this test doesn't guarantee the quality PRNG in general or in any other application. An advice that was made by Ferrenberg, Landau & Wong long time ago is still valid:

> *"…a specific algorithm must be tested together with the random number generator being used regardless of the tests which the generator has passed."* [25]

Recommendations to practitioners regarding PRNGs sometimes sound a bit relaxing. Dupree and Fraley state:

> *"Fortunately, imperfections of the random number generators are frequently minor and can usually be tolerated by Monte Carlo practitioners."* [49]

Halton states:

> *"Since we can neither prove that any particular process or device is a priori random, nor test its output exhaustively, the search for randomness is evidently futile. This would be very discouraging, were it not for the fact, when "random numbers" are used in practice, we generally require only a few of the properties of randomness, and all others are immaterial."* [50]

We believe that the above quotations do not declare randomness tests irrelevant. They talk to the point that it is the requirements of a specific simulation task that to the large extent determine whether a particular PRNG can be considered suitable.

Wrapping up a quick overview of the general PRNG development status, we should note that despite extensive research in the field of random number generation and serious success gained over the two – three recent decades, ideal PRNG is still a matter of the future. There's no theoretical solution to determine suitability of PRNG for a specific application. In each case, a combination of PRNG (or several generator options), MC algorithm and computer hardware/software should be tested and matched for an efficient result [5, p. 33; 92]. Anyway, as a starting point for testing any combination of PRNG, and computer environment, it's much better to use a validated PRNG (especially, if they are available) than a generator with undocumented qualities.

**Excel 2010 Random Number Generators.** Table 1 combines information on several types of Excel's PRNGs. All generators for the versions up to and including Excel 2007 were known to be inadequate because of small period length and unclear algorithms [52, 77, 87].

There are three PRNGs in Excel 2010 [87, 89]:

– The generator called upon by function RAND.

– The generator in the Statistical Toolpak add-in.

– The generator called upon by VBA RND function.

Guy Melard shows that PRNGs of the Statistical Toolpak and VBA have not been changed from previous versions [87]. So, these generators are still inadequate for simulations and shouldn't be used.

Microsoft indicates that RAND generator has been improved for Excel 2010 [88]. Guy Melard conducted a test of the new PRNG with a modified (simplified) TestU1 [52], and found that the new PRNG has satisfactory statistical qualities [87].

Some authors state that a new RAND PRNG employs Mersenne Twister algorithm [84, 87]. This statement is based on a reference to the Microsoft blog [88]. When we accessed this page, there were no specifics regarding the type of the new RPNG. Microsoft knowledge article on the RAND function still (as of March 2015) indicates that Excel 2010 RAND function is based on the Wichmann-Hill algorithm [90]. We agree with Stephan Levy that, despite acknowledging improved quality of the RAND PRNG, there's much uncertainty about the RAND function: is it really Mersenne Twister algorithm, was it implemented correctly, what's the period length? [89].

**Table 1: Suitability of Excel functions by version**

| **Function** | **Excel 97** | **Excel 2003** | **Excel 2007** | **Excel 2010** |
|---|---|---|---|---|
| **Random number generator** | | | | |
| RAND | Inadequate [72] | Inadequate [73, 77, 78] | Inadequate [77, 78] | Improved [87]. Undocumented [89]. |
| RND | | | Inadequate | Unchanged [87] |
| Analysis Toolpak (ATP) | | Inadequate [73, 77, 78] | Inadequate [79] | Unchanged [87] |
| **Statistical distributions** | Inadequate [72] | | | |
| Standard normal | Erroneous [73] | Fixed for function call NORMSINV(RAND). Not fixed for analysis toolpak (ATP) [73] | | No errors found for Norm.S.Dist and Norm.Dist [86]. Improved [87]. |
| Inverse standard normal | Weak [73] | Fixed. Exact [73] | Inadequate [84] | Exact result [84]. Functions Norm.Inv and Norm.S.Inv can give errors for small values of probabilities [86]. |
| Poisson | Bugs [74] | Bugs [74] | Inadequate [80, 91] | Performs perfectly along with R and SAS [84]. No errors found for Poisson.Dist when computing point probabilities [86]. Improved [87]. |
| Binomial | Bugs [74] | Bugs [74] | Inadequate [80, 91] | 100% correct [84]. No errors found for Binomial.Dist [86]. Improved [87]. |
| Inverse binomial | | | | No errors found for Binom.Inv [86]. |
| Gamma | Bugs [74] | Questionable [74] | Inadequate [84]. | Acceptable [84]. Gamma.Dist can give errors |

| | | | Questionable [91] | for small values of probabilities [86]. Improved [87]. |
|---|---|---|---|---|
| Inverse Gamma | | | | Gamma.Inv can give errors for large values of the parameter *a*. Gamma.Inv and Gamma.Dist are not consistent [86]. |
| Beta | | | Inadequate [84] | Acceptable [84]. No errors found in Beta.Dist for density function. Left tail probabilities can be chaotic [86]. Improved [87]. |
| Inverse Beta | Bugs [74] | Questionable [74] | Inadequate [80]. Unreliable [91] | Beta.Inv can give errors for certain parameters [86]. |
| Student's *t* | | | | For T.Dist, T.Dist.2T, T.Dist.RT result is not always correct [86]. Improved [87]. |
| Inverse student's *t* | | | Inadequate [80, 84, 91] | Acceptable [84]. For T.Inv and T.Inv.2T results can be completely wrong [86]. |
| F | | | | F.Dist and F.Dist.RT – no errors found for density function, even for small probabilities [86]. Improved [87]. |
| Inverse F | | | Inadequate [80, 84, 91] | Acceptable [84]. No errors found with F.Inv and F.Inv.RT [86]. |
| Chi-square | | | | Chisq.Dist, Chisq.Dist.Rt. No errors found for density function. Results can be wrong if probabilities are small [86] Improved [87]. |
| Inverse Chi-square | | | Inadequate [84]. Unreliable [91] | Questionable. One of 15 distributions miscalculated [84]. No errors found in Chisq.Inv. Chisq.Inv.RT and Chisq.Dist.RT are not |

| | | | | consistent [86]. |
|---|---|---|---|---|
| Hypergeometric | | | Inadequate [84] | 100% correct [84]. No errors found for Hypgeom.Dist [86]. Improved [87]. |
| **Univariate summary statistics** (mean, the sample standard deviation, the correlation coefficient, mode, median, maximum, minimum). Excel commands for computing these quantities are: 'Average', 'Stdev', 'Pearson', 'Median', 'Mode', 'Max', 'Min'. | Inadequate [72] | Acceptable [73]. Standard deviation problems corrected [75] | Basic descriptive statistics can be used with confidence [81, 84] | Acceptable. Means accurate to 15 significant digits [84]. |
| **Regression** | | | | |
| Linear regression | | Acceptable [73]. Improved. Performance on computing R-squares and beta coefficients is similar to that of SAS [75] | Acceptable. Performs better than previous version [83, 91]. Except no warning of detecting perfectly collinear data [79] | Acceptable. On some data sets inferior to previous version [84]. |
| Nonlinear regression | | Unacceptable [73, 75]. Accuracy is far below major statistical packages [75] | No changes. Inadequate [79]. Results are better than in other studies [83]. | No changes [87] |
| **Other** | | | | |

| Functions | | | | |
|---|---|---|---|---|
| Exponential smoothing | | | Inadequate [79] | |
| LOGEST | | | Inadequate [79] | |
| GROWTH | | | Inadequate [79] | |
| Trendline (ATP) | | Inadequate [85] | Inadequate [79, 85] | |

### 2.2. Excel 2010 accuracy of the statistical distributions

Table 1 combines results of testing of statistical distributions for several generations of Excel. Inaccuracies of statistical distributions in Excel have been well documented for the versions up to and including Excel 2007. Many commonly used distributions (e.g. Poisson, Beta, Gamma, Binomial, etc.) were considered inadequate [74, 80, 84].

Excel 2010 marks a step in a long-awaited direction. Two papers were published with test results of the Excel 2010 statistical distributions by the authors known for testing and critical discussions of the Excel's prior versions: Kellie Keeling and Robert Pavur [84] and Leo Knusel [86]. Both articles confirm serious improvements and elimination of many errors identified before. Some distributions were found completely error-free in both articles: e.g. Poisson, Binomial, Hypergeometric. Several distributions were found inaccurate for extremely small values of probabilities, e.g. Gamma, inverse normal [86]. Some results need to be verified for the testing conditions and parameters. For the inverse *t*-distribution, article [84] reports perfect accuracy (same as R statistical package), but article [86] admits that the results could be completely wrong for this function. Also, according to [84], all tested distributions are very accurate, except inverse Chi-square, but no errors in this distribution were found in [86]. Most recent contribution by Guy Melard, confirms improvements (compared to prior Excel versions) with most distributions [87].

### 2.3. Excel 2010 univariate summary statistics

Univariate summary statistics include mean, the sample standard deviation, the correlation coefficient, mode, median, maximum, minimum. Excel commands for computing these quantities are: 'Average', 'Stdev', 'Pearson', 'Median', 'Mode', 'Max', 'Min'. Arguably, these functions are used in any MC simulation study. Tests show that in Excel 2010 basic descriptive statistics can be used with confidence. Means are accurate to 15 significant digits [81, 84].

### 2.4. Excel 2010 regression functions

Calculation of the linear regression was deemed acceptable since Excel 2007 version [83, 91]. Excel 2010 also demonstrates acceptable performance, although its linear regression accuracy on some data sets is inferior to the previous version [84]. Non linear regression of Excel 2007 was very weak [79, 83]. No changes were made in Excel 2010 [87].

### 2.5. Excel 2013

Excel 2013 has 104 functions in the statistical category [98] including six new functions (BINOM.DIST.RANGE, GAMMA, GAUSS, PERMUTATIONA, PHI, SKEW.P [99]). The literature search didn't retrieve academic papers on testing statistical functions in Excel 2013 (except an article on Excel 2013 in the cloud [93], see below). So there's no third-party information on the quality of the Excel 2913 statistical functions.

One of the types of Excel services offered by Microsoft is Office Online (formerly Office Web Apps). The most current version of the application is offered through the online service. McCullough and Yalta examined this Excel service [93]. Various Excel Web App functions in the cloud have demonstrated mixed levels of accuracies. However, the accuracy seems not to be the major consideration in this case. The testing has revealed a portability issue: the same spreadsheet opened in the cloud and on the desktop may give different results [93]. Even bigger concern is that the cloud option leaves researchers with little to no information about the hardware and software that is used by the service provider, not to mention that the elements of the infrastructure could be changed (with the best intention of improving services) at any time without a notice [93]. These concerns don't allow us to recommend the use of Excel cloud services (at least at this time).

## 3. Validation and Verification of MC Simulation Models and Sources of Error

### 3.1. Sources of error

By the very nature of the Monte Carlo method, it deals with uncertainties and errors. Despite variousness of the uncertainties and errors, they could be categorized into several groups according to the phases of modelling and simulation (described in the next subsection) [55, 56, 66]. It should be noted that numerical amounts of errors mentioned below apply only to the specific simulations they represent and may not be generalized on any other situations. These data demonstrate potential levels of errors from different sources.

**Table 2: Sources of error in MC simulation**

| Sources of Error |
| --- |
| **Simulation project conception** |
| Errors due to unclear definition of simulation project objectives, in-scope system parameters, assessment criteria and required accuracies. |
| Errors due to linguistic uncertainty, e.g. vagueness, ambiguity [61]. |

**Input data analysis**

Errors due to imperfections of input data, e.g. errors, inaccuracies of collected data characterizing modelled systems or processes [55].

Robinson presented a simulation case of a simple queue line in a bank [55]. The simulation has shown that underestimation or overestimation of the service time by 10% led to the significant underestimation of 30% or overestimation of 60%.

Errors due to inadequate sample size of collected data characterizing modelled systems or processes [55].

Errors due to imperfect descriptions of the input data with selected or fitted probability distributions.

Errors due to incorrect assumptions about input data for the situations in which real systems or processes are not available for empirical data collection.

Errors due to incorrect modelling of randomness, e.g. selecting probability distributions that incorrectly represent randomness of the real system or process [56].

Law and McComas presented a simple simulation case of a single machine tool system with exponential interarrival times [56]. They have shown that incorrect selection of normal or lognormal input distributions (instead of a correct distribution which is Weibull in this case) led to large output errors of estimated average delays of 39% and 65% respectively.

Errors due to incorrect selection of the parameters of probability distributions or replacing distributions by their means [56].

**Conceptual modelling**

Errors of conceptual modelling arising due to incomplete knowledge and understanding of the modelled system or simplification of the real system behaviour, e.g. ignoring correlations and co-variance in input distributions [55, 58].

Errors due to overcomplication of the conceptual model with multiple estimated parameters (parametric uncertainty) [59, 60].

Errors of mathematical modelling due to approximations and simplifications [66].

**Converting conceptual to computer model**

Errors of converting a conceptual model into computer model due to misinterpretation of model description or incorrect application of mathematical theory [5, 55, 62].

Errors due to systematic errors due to programming errors.

| | |
|---|---|
| | Errors due to misuse of simulation software by underqualified users [56], e.g. using default settings which are unacceptable for the specific problem. |
| | Errors due to typing errors [62] |
| **Experimentation** | |
| | Errors due to ignoring the model's initial transient period or, more generally, incorrect initial conditions [55]. |
| | In the simulation of a simple queue line in a bank, presented by Robinson, it was shown that the output error of ignoring initial transient period was 4.57% [55]. |
| | Errors due to insufficient number of replications (simulations runs) [5, 55, 65]. |
| | Errors due to insufficient searching of the solution space [55], e.g. system's parameters and influencing factors are changed in limited ranges and don't reveal full model's behaviour. |
| **MC software incurred errors (part of experimentation)** | |
| | Errors due to low quality of the PRNG. |
| | Errors due to imperfect transformation of the PRNG to other probability distributions. |
| | Errors due to numerical approximations [58]. |
| | Errors due to limited resolution in space and time (discretization) [58]. |
| | Errors due to bugs in software [58] |
| | Errors due to round-off errors induced by simulation computer hardware and operating system due to limited word length [5, 67, 71]. |
| | Errors due to truncation errors induced by simulation computer hardware and operating system due to limited word length [5, 67, 71]. |
| | Errors due to algorithmic inefficiencies, e.g. using Newton method for nonlinear equations [67, 68]. |
| **Simulation output data analysis** | |
| | Errors due to incorrect estimation of the output probability distributions of variables of interest and their parameters (e.g. mean and variance) [64]. |
| | Errors induced by visualization of results [66]. |

Each MC simulation research effort is unique and depending on multiple factors (such as the purpose of the simulation, size and complexity of the modelled system or process, application area, etc.) it will experience different types and levels of uncertainties and errors. Complex combination of these errors and uncertainties will propagate through the model and impact the accuracy of the estimated output parameters.

### 3.2. Verification and Validation

Some of the errors and uncertainties outlined above can be decreased or completely eliminated (e.g. programming bugs). Others are inherent and are not reducible. Verification and validation processes are part of the overall simulation model development efforts undertaken in order to assure that the models are built according to the user objectives, correctly represent real world systems or processes of the application domain and produce credible and reliable results. For the purpose of this paper, we are interested in certain aspects of verification and validation which are related to the simulation computerized tools and programming:

> *"Operational **validation** is defined as determining that the model's output behaviour has a satisfactory range of accuracy for the model's intended purpose over the domain of the model's intended applicability."* [63]

> *"Computerized model **verification** is defined as assuring that the computer programming and implementation of the conceptual model are correct."* [63]

It is important to note that the process of MC simulation validation is not targeted to achieving maximum accuracy. Required accuracy is determined by the purpose of the simulation to satisfy the objectives of the decisions to be made based on the simulation results. It is intuitively understandable, that different simulation scenarios will require different model accuracy. For example, simulation that reproduces in detail a history of a stochastic movement of a high-energy particle including its trajectory and interactions with surrounding particles will most likely require higher accuracy than a model of a waiting line to a bank teller. A computerized tool or a program that is not suitable for the physics scenario due to lack of accuracy may perfectly fit the needs of the operational bank evaluation.

A crucial point here is that a discussion of the accuracy of a simulation tool (and hence its suitability for MC simulation) cannot be meaningful without relation to a specific situation (task, purpose, users, decisions to be made), which determines the requirements to the acceptable levels of the simulation output accuracies.

## 4. Numerical Experiment

Numerical experiment involved two computers. Their characteristics are shown in Table 3. Computer 1 (C1) had Windows 7 and Excel 2010, and computer 2 (C2) had Windows 8.1 and Excel 2013. Both Excel applications are 32-bit installations.

**Table 3: Computers used in numerical experiments**

|  | **Computer 1** | **Computer 2** |
|---|---|---|
| Operating System | Windows 7 Professional | Windows 8.1 |
| Service Pack | Service Pack 1 |  |
| Word Length, bit | 32 | 64 |
| Processor | Intel, Core i7-3520M | Intel, Core i5-3210M |
| CPU Speed, GHz | 2.90 | 2.50 |
| Memory Installed, GB | 4.0 | 6.0 |
| Memory Usable, GB | 2.96 | 5.87 |

On both computers, the following three PRNGs were used: two Excel built-in functions RND, RAND and Mersenne Twister PRNG. RAND function (as it is not available in VBA directly) was called through Evaluate("=Rand()"). MTwister is a translation by Jerry Wang of the C-program Mersenne Twister MT19937 into an Excel VBA module "MersenneTwisterVBAModule" (as of October 2014). This module is freely available at the website of the Mersenne Twister algorithm creators Takuji Nishimura and Makoto Matsumoto at the University of Hiroshima [100] and compatible with Excel 2010 and Excel 2013.

**Visual testing.** Fig 1 through 3 demonstrate visual distribution of 1,000 random numbers generated by RND, RAND and MTwister PRNGs, respectively. None of the generators reveals any irregularities or repeating patterns.

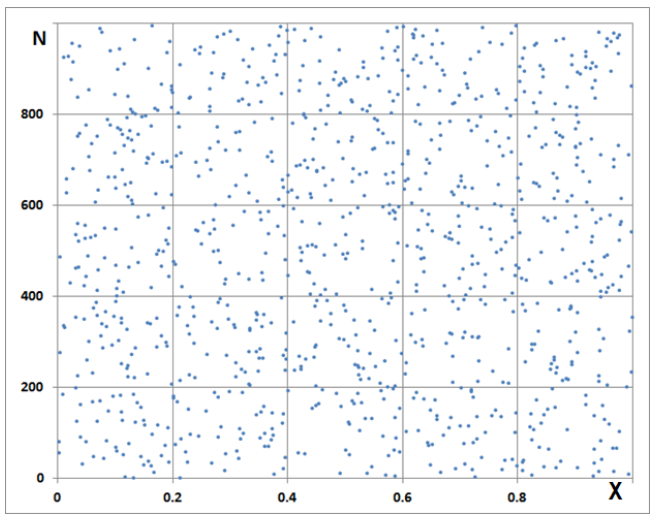

**Fig 1: Visual representation of 1,000 numbers generated by RND PRNG (C1 Excel 2010)**

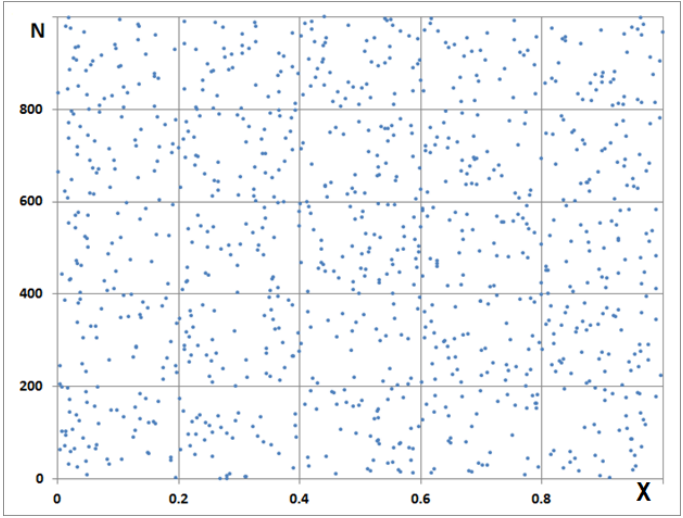

**Fig 2: Visual representation of 1,000 numbers generated by RAND PRNG (C1 Excel 2010)**

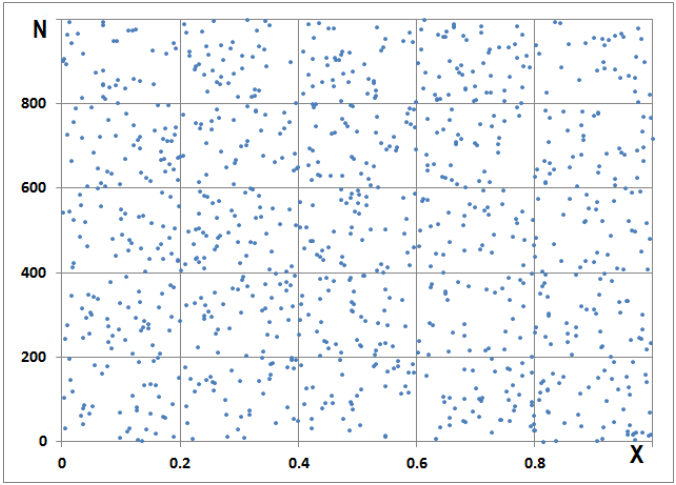

**Fig 3: Visual representation of 1,000 numbers generated by MTwister PRNG (C1 Excel 2010)**

Histograms on Fig 4 and Fig 5 show distribution of random numbers generated by RND and MTwister, respectively. These graphs allow for approximate comparison of the PRNGs' uniformity. The MTwister demonstrates better uniformity.

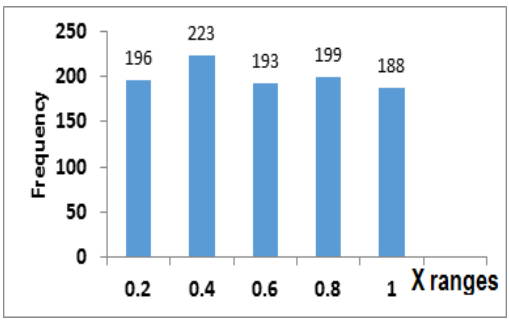 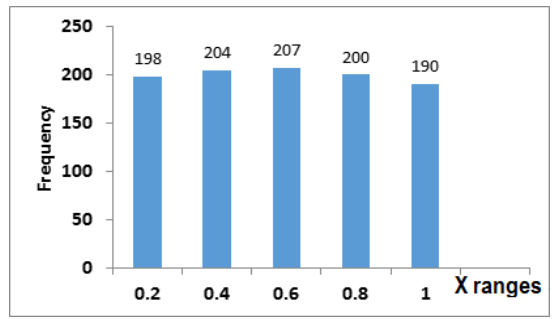

**Fig 4: Histogram for RND PRNG**         **Fig 5: Histogram for MTwister PRNG**

**Chi-square Test.** Ten thousand random numbers were used from MTwister PRNG (from 1,000,001[st] to 1,010,000[th] with a default seed 5489). The Chi-square goodness-of-fit uniformity test followed a procedure described in [101]. Calculations and results are shown in Fig. 6.

| | O | P | Q | R | S | T | U |
|---|---|---|---|---|---|---|---|
| 1 | Bins | Freq | Expected | Chi SQ | | | |
| 2 | 0 | 0 | | | | | |
| 3 | 0.1 | 982 | 1,000 | 0.3 | =(Q3-P3)^2/Q3 | | |
| 4 | 0.2 | 1030 | 1,000 | 0.9 | | | |
| 5 | 0.3 | 1030 | 1,000 | 0.9 | C2 statistics= | 8.7 | =R13 |
| 6 | 0.4 | 959 | 1,000 | 1.7 | degFreedom= | 9 | =COUNT(R3:R13)-1 |
| 7 | 0.5 | 948 | 1,000 | 2.7 | alpha= | 0.05 | |
| 8 | 0.6 | 1025 | 1,000 | 0.6 | critVal= | 16.919 | =CHIINV(T7,T6) |
| 9 | 0.7 | 983 | 1,000 | 0.3 | P-value | 0.46541 | =CHIDIST(T5,T6) |
| 10 | 0.8 | 1002 | 1,000 | 0.0 | H0 - Data is uniformly distributed | | |
| 11 | 0.9 | 1036 | 1,000 | 1.3 | H1 - Data is not uniformly distributed | | |
| 12 | 1 | 1005 | 1,000 | 0.0 | | Accept | Null hypothisis |
| 13 | | | | 8.7 | =SUM(R3:R12) | =IF(T5>T8,"Reject","Accept") | |

**Fig 6: Spreadsheet calculations and results of the Chi-squire test of MTwister PRNG**

The results confirm that the MTwister PRMG implemented in VBA generates uniformly distributed data. Statistics equals 8.7 which is way below 16.9 that allows accepting the hypothesis with 95% confidence.

**Computational speed testing.** Each speed test involved recording time for generation and writing to a spreadsheet 1,000,000 random numbers. The results are presented in Table 2.

**Table 4: Computational speed testing (time in seconds)**

| | RND | RAND | MTwister |
|---|---|---|---|
| C1 | 4 | 20 | 10 |
| C2 | 25 | 46 | 12 |
| Average | 14.5 | 33 | 11 |

Computational speeds of the RND and MTwister are approximately at the same level. Slower speed of the RAND generation can be explained by the fact that it was called through the EVALUATE function.

## 5. Discussion

As it was mentioned, MC is used in multiple industries and knowledge fields to solve a wide variety of problems. Real-world problems may require specific mathematical apparatus to model them and, hence, numerous types of computer functions to support simulation of the models. Functionality of the MC simulation software can be subdivided into two categories. The first one, which we call *Core*, includes data input and results output facilities, programming environment, PRNG and functions for calculating summary statistics. All of the core components are arguably used in any simulation project forming the minimum scope of MC software functionality. Also, for many applications these components can be the only ones necessary and sufficient to complete the project. The second category, *Incidental*, includes statistical distributions, linear and non-linear regression and other statistical, engineering and business functions. Commonly, only some of the incidental components are required for a given project. Incidental components cannot be used by themselves to form a complete simulation project: they are used in addition to the core components. Core and incidental components of the MC simulation tools are shown in the left side of the Figure 7. A component could represent either one function (e.g. PRNG) or a group of many functions with similar purpose or application (e.g. a group of statistical functions include over 100 individual Excel functions).

| Components | | Excel 2010 | Excel 2013 |
|---|---|---|---|
| Core | Data Input & Output Facilities | + No concerns with spreadsheets | + No concerns with spreadsheets |
| | Programming Environment | + No concerns with VBA development environment | + No concerns with VBA development environment |
| | Pseudo-random Number Generator | Use with external VBA PRNG; -- Do not use with built-in PRNGs: RAND, RND | Use with external VBA PRNG; -- Do not use with built-in PRNGs: RAND, RND |
| | Summary Statistics | + No concerns with summary statistics, e.g. AVERAGE, STDEV | + No concerns with summary statistics, e.g. AVERAGE, STDEV |
| Incidental | Statistical Distributions | + / ~ / -- | + / ~ / -- |
| | Linear Regression | + | + |
| | Non-Linear Regression | -- | -- |
| | Other Functions (statistical, engineering, business) | + / ~ / -- | + / ~ / -- |

Legend: **+** Use with confidence; **~** Test before use; **--** Do not use
**+ / ~ / --** Mix of functions with different levels of usability

**Fig 7: Components of the MC simulation tools and suitability of Excel functions**

The right two columns of Figure 7 demonstrate how Excel 2010 and Excel 2013, respectively, qualify in each of the MC simulation components. At a high level, qualifying characteristics of suitability can be expressed by four categories. The first is *Use with confidence*. It means that the component has been tested and displayed positive results. If there are more than one function in the component, it means that *all* functions were tested and proved to be good. The second is *Do not use*. It also means that the results of the tests are available but they were negative. This category also can include a component with several functions - all of them should not be used. The third category is *Test before use*. It applies to the components that either were not tested in the third-party reviews or the results were mixed. These components require careful assessment/testing before they could be included in the MC models. This category is arguably the most numerous: relatively small number of Excel functions (compared to the total number of functions available) received third-party testing. Finally, the fourth category represents components with multiple functions with mixed levels of usability.

Qualifying characteristics of Excel 2010 and Excel 2013 are shown in the two right columns of Figure 7 and are based on the published results of the Excel functions testing (combined in Table 1), and analysis and numerical experiment of this study. For both Excel 2010 and Excel 2013, most core components are displaying good results, i.e. input and output with spreadsheets, VBA development environment and summary statistics can be used with confidence in MC simulations. The only exception in the core components is Excel PRNG and it needs more comments.

As it has been demonstrated in Sections 2 and 5, Excel built-in PRNGs have serious deficiencies. VBA generator RND has been known to have poor statistical qualities. PRNG RAND, although it has been updated in Excel 2010, has not been documented – so its period length is unclear. Periodicity requirement is determined by the number of calls that will be made during one simulation. The weaker requirement is that the period of length cannot be less than number of calls to PRNG in one simulation to avoid repetition of the same random numbers. The stronger requirement should also take into account a risk of long-range correlations in number sequences. To alleviate this risk it is recommended to use only a small part of the period – usually no more than 10% of the numbers in the period [50], or even that the period of a PRNG should be at least 200 times greater than the square of the number of pseudo-random numbers needed [SQRT (length/200)] [89, 103]. In some situations, the requirement may be more stringent. For example, L'Ecuyer and Simard have shown that for two dimensional uniformity of pseudo-random numbers generated by a PRNG with cycle length L, one should not use sample sizes more than approximately equal to the cubic root of the period length (or modulus) [57]. Not knowing exact period length of the RAND PRNG puts researchers in an uncertain situation regarding suitability of the software tool. Also, RAND is not a built-in VBA function and calling it through EVALUATE function slows down simulations. Another serious issue with Excel built-in PRNGs, which is not commonly mentioned, is that they use current machine time as a seed number. The researchers can't exercise seed control which means that results of the simulations become irreproducible [77, 89]. Mandatory reproducibility of research is an upcoming ethical issue and requirement of the contemporary science. In general, reproducibility is a very broad notion that includes documenting and archiving input and output data, description of the model and code used, etc. [102]. For the MC simulation field, inability to set and document a PRNG seed number makes any other efforts aimed at reproducibility of research impossible and senseless. So, current PRNGs add an ethical problem to the acknowledged technical issues of Excel. Aside from purely ethical implications, lack of seed control may also have tangible consequences, if research results become a matter of litigation [89] or audit under Sarbanes-Oxley Act. Overall recommendation is not to use built-in Excel PRNGs for MC simulation. Fortunately, this restriction can be easily overcome with a use of external to Excel PRNG implemented as a VBA module. In this study, we used Mersenne Twister MT19937 PRNG as an Excel VBA module "MersenneTwisterVBAModule" available at the website of the Mersenne Twister home page at the University of Hiroshima [100]. As a Mersenne Twister type generator, it has a period of length that can satisfy most random number extensive applications – $2^{19937}-1$. This PRNG demonstrated seed control, easy integration with a VBA program and random numbers generation speeds comparable with the RND built-in function. We believe that this PRNG can be used in MC simulation to replace Excel built-in PRNGs and eliminate their problems. The same website provides another VBA version of the Mersenne Twister PRNG [100]. It should be noted that literature search didn't retrieve any academic papers with extensive tests of the VBA PRNGs, so in making recommendation to use these PRNGs we rely on the reputation of the Mersenne Twister algorithm creators offering these programs on their website. We are not questioning the skills of the people who translated Mersenne Twister programs from C into VBA, however, there could be some subtle specifics in the properties of both languages that may incur risks of "something being lost in translation". Thorough tests of the randomness qualities of the VBA PRNGs using the TestU01 [52] or Diehard tests [53] are highly desirable.

To complete a discussion of the core components, we can reiterate that input and output data with spreadsheets, VBA development environment and summary statistics can be used with confidence in MC simulations. Complemented with an external PRNG added as a VBA module (e.g. "MersenneTwisterVBAModule" or similar), the core components of the Excel 2010 and Excel 2013 can serve as a solid simulation framework. Based on our definition of the core components, it is clear that this framework can be used to implement a broad variety of MC simulation projects. Actually, any other elements of the simulation model (if necessary) could be just programmed with VBA. Whether this approach is feasible or makes sense from a workload point of view depends on a specific project.

Incidental components represent several groups with multiple and disparate functions (even within a single group) (see Fig 7). Linear regression and non-linear regression are easier to qualify: linear regression can be used without concerns and non-linear regression should not be used (see Table 1). Suitability of the functions which are included in the statistical distributions component vary from those which were tested as completely error-free and can be used with confidence (e.g. Normal, Poisson, Binomial distributions), to those with mixed test results (e.g. inverse t-distribution). The same refers to the other statistical, engineering or business functions component. In each case of considering functions from incidental components due diligence should be exercised by analysing results of the third parties' reviews (see Table 1) or conducting additional tests of candidate functions. It should be noted that special caution should be taken with Excel 2013 as there is lack of available reviews (at least at present). Newly added functions should be tested before they can be recommended for use in simulation models. Even improved functions (which tested positive in Excel 2010) should be re-tested, as the vendor has some history of making new mistakes while correcting previous ones.

Certain miscommunication can be found in the literature. Many papers analyze statistical software and requirements to this type of solutions. Because MC simulation is based on statistical methods, there's a temptation to generalize and directly apply all requirements to the statistical software to the MC simulation software. This may not be always a correct approach. A conclusion of this study is that "Excel 2010 is a strong MC simulation tool…" with certain conditions and limitations. It may be perceived incompatible with a widely publicized notion that "Excel should not be used as a statistical software package…' – after all we are talking about the same software application with the same set of statistical capabilities. However, there is no contradiction. Conclusions reflect the business needs which transform into varying assessment expectations, requirements and assumptions.

First, when Excel is assessed as a statistical software package, the expectation is that the *accuracy of the results must be perfect* (usually, exact to 15 significant digits). Because this is not always achieved – the tool is deemed inappropriate. When Excel is assessed as a MC simulation tool, the *accuracy must be good enough* to satisfy the objectives of the decisions to be made based on the simulation results. The requirements to accuracy may vary from very high to rather relaxed based on the needs of a specific simulation project, and Excel may or may not meet these requirements.

Second, when Excel is assessed as a statistical software package, another expectation is that *all statistical functions* must work perfectly and nomenclature of the functions must match variety of functions available in other statistical software. Because this is not always achieved – the tool is considered inappropriate. When Excel is assessed as a MC simulation tool, the expectation is that the *core components* must provide a reliable simulation framework while the use of the incidental components has to be determined through the validation and verification process. If certain incidental functions are not available or not performing well enough, VBA development environment may be used to program missing components.

Using an analogy, statistical software packages evaluators are working on competitive Formula 1 bolides and their requirements may or may not be always applicable to a case of selecting a family car.

Also, it should be noted that Excel 2010 and Excel 2013 accumulated many improvements (may be not as quickly and thoroughly as the users would need or expect) and certain critical statements regarding the statistical capabilities of this application, which still could be come across in the literature, do not apply any more.

The importance of this work is in offering an objective view of the recent Excel versions through the lens of the MC simulation needs. Practical implication of the paper is in showing a pragmatic approach to using Excel's strengths and avoiding mistakes in simulation projects. Also, we believe that the overall cast of doubt that tends to overshadow all MC simulations using Excel will be lifted (at least from implementations that use Excel with caution and follow the best V&V practices).

Future research will focus on in-depth testing of Mersenne Twister VBA PRNGs and accuracy assessments of the Excel 2013 statistical distributions.

## 6. Conclusions

1. Microsoft Excel (versions 2010 and 2013) is a strong Monte Carlo simulation application. It offers a solid framework of core simulation components including spreadsheets for data input and output, VBA development environment and summary statistics functions, which have been tested to provide reliable performance. Researchers should complement this framework with an external high-quality PRNG added as a VBA module (e.g.

"MersenneTwisterVBAModule" or similar). Even the Excel framework of core simulation components alone can be used to implement a broad variety of MC simulation projects.

2. Excel is also offering a large and diverse category of incidental simulation components that includes statistical distributions, linear and non-linear regression and other statistical, engineering and business functions. By using these components, development of the simulation models can be expedited. Suitability of the functions in this category vary from completely error-free to those with mixed test results or those with unknown qualities. Due diligence should be exercised when considering functions from the incidental components for the project: conduct statistical tests of candidate functions and/or analyze results of the third-party reviews (e.g. see Table 1).

3. The general suitability of Excel 2010 for MC simulation, stated in the first point of conclusions, does not imply that Excel is appropriate for any simulation task. Each simulation project is unique and should be based on the best verification and validation practices, which involve, among other steps, determining the acceptable range of accuracy of estimated output parameters to satisfy the needs of decision makers; identifying Excel functions that are required for the model and assessing their statistical qualities and related errors; identifying all model errors and uncertainties and evaluating their propagation through the model and impact on the overall accuracy (including the contribution of the Excel-induced errors). Only after comparing acceptable and expected errors and performing test runs of the model on the project-specific computer environment, researcher can make an informed decision on the suitability of Excel for a given simulation project.

**Acknowledgements**

The author is grateful to Natalia Botchkareva, a financial analyst with expert knowledge and skills in Excel, for constant support and advice.

The author is thankful to an anonymous referee who rejected author's previous paper in another journal due to the use of Excel spreadsheet MC simulation – dimmed inappropriate. That motivated work on this article.

The views, opinions and conclusions expressed in this document are those of the author alone and do not necessarily represent the views of the Ontario Ministry of Health and Long-Term Care or any other organizations the author is affiliated with.